%% file: main.tex
\documentclass[fleqn,10pt]{wlscirep}
\usepackage[utf8]{inputenc}
\usepackage[T1]{fontenc}
\usepackage{lineno}
\usepackage{adjustbox}
\usepackage{array}
\usepackage{multirow}
\usepackage{ltablex}

\title{Russian Financial Statements Database: A firm-level collection of the universe of financial statements}

\author[1]{Sergey Bondarkov}
\author[1]{Viktor Ledenev}
\author[1]{Dmitry Skougarevskiy}
\affil[1]{European University at Saint Petersburg, The Institute for the Rule of Law, Saint Petersburg, 191187, Russia}

\affil[*]{Corresponding author: Viktor Ledenev (\href{mailto:vledenev@eu.spb.ru}{vledenev@eu.spb.ru})}

\begin{abstract}
The Russian Financial Statements Database (RFSD) is an open, harmonized collection of annual unconsolidated financial statements of the universe of Russian firms in 2011--2023. It is the first open data set with information on every active firm in the country, including non-filing firms. With 56.6 million geolocated firm-year observations gathered from two official sources, the RFSD features multiple end-user quality-of-life improvements such as data imputation, statement articulation, harmonization across data providers and formats, and data enrichment. Extensive internal and external validation shows that most statements articulate well while their aggregates display higher correlation with the regional GDP than the previous gridded GDP data products. We also examine the direction and magnitude of the reporting bias by comparing the universe of firms that are required to file with the actual filers. The RFSD can be used in various economic applications as diverse as calibration of micro-founded models, estimation of markups and productivity, or assessing industry organization and market power. 
\end{abstract}
\begin{document}

\flushbottom
\maketitle

\thispagestyle{empty}

\section*{Background \& Summary}

Financial statements are the main source of publicly available information about firms. Any economic analysis based on national or regional aggregates may overlook the underlying heterogeneity of individual firms\cite{bartelsman2000understanding, melitz2003impact, mian2010great}. While many firm- or plant-level surveys have become freely available since the 1990s\cite{bernard2007firms}, including from developing countries\cite{roberts1997decision, levinsohn2003estimating}, country-wide open firm-level datasets are still rare. Instead, scholars resort to commercial databases maintained by business information publishers: Moody's ORBIS, S\&P's Compustat, or Refinitiv's Worldscope. However, firm coverage and representativeness of the commercial sources is often low\cite{liu2020}, especially for developing countries and emerging markets\cite{nam2018, mcguire2016}. Commercial databases are also notorious for the ambiguity in the definition of the unit of analysis, imprecise or missing values, selection and survival bias, reporting lags, lack of transparency, unidentifiable coding errors, steep learning curve, and restrictive access costs\cite{bajgar2020, chychyla2015, dai2012, goswami2019}. These problems can materially affect study results\cite{lara2006, mcguire2016, nam2018}. Some researchers choose to construct derivative versions of the commercial data sets\cite{ribeiro2010, kalemli2024construct, almunia2018evaluating} to overcome these problems while others create their own databases from public or administrative sources\cite{nagel2001,lopez2015assessing,berlingieri2017multiprod,wahlstrom2022financial}.

Here we present the Russian Financial Statements Database (RFSD)~--- an open, harmonized dataset with the universe of annual financial statements filed by Russian firms in 2011--2023. This dataset with 56.6 million firm-year observations is built from the administrative data and features a number of end-user quality-of-life improvements such as data imputation, statement articulation, and harmonization across data providers and formats. 

Extant literature relying on Russian firm-level data studies corporate governance \cite{schweiger2013,muravyev2017, sprenger2022,garanina2021}, financial transparency and reporting\cite{bagaeva2008a,braguinsky2015}, allocative efficiency\cite{cusolito2020,kaukin2023allocation,abramov2023tfp, bessonova2023budget}, firm entry\cite{bruno2013}, corruption or embezzlement\cite{mironov2016,mironov2013}, regulatory capture\cite{slinko2005}, access to credit\cite{shvets2013judicial}. Firm-level data is also harnessed by researchers in government\cite{mogilat2024} and international organisations\cite{schweiger2013, wildnerova2020}. To date, these and other firm-level studies of the Russian economy have relied almost exclusively on commercial databases: Moody's Orbis (and its component Ruslana) or Interfax's SPARK\cite{goswami2019}. Problems of data imbalance, under-representation of small and medium-sized enterprises, and missing values loom large in these sources\cite{abramov2023tfp, bessonova2023budget,zhemkova2023allocation}. For instance, Moody's Ruslana represents about 10\% of total wage employment and poorly covers some industries such as real estate or educational services\cite{goswami2019}.

The RFSD responds to the growing demand for open and reliable firm-level data on the Russian economy. Taking careful account of the changes in reporting standards, forms, and rules, we source administrative data on financial statements and unify the information from company balance sheets, income statements, cash flow statements, etc. into a single flat table. We then use the following-year statements to impute missing values in the current-year statements and reconstruct the data wherever possible. We further correct the errors in the financial statements by restating them in accordance with the applicable accounting rules.

Crucially and in contrast to other sources of firm-level information, we enrich the data in the RFSD with the information on non-filers~--- that is, companies that were legally required to submit their statements and did not benefit from any exemptions but failed to do so. To do this, we gather the information on the universe of legal entities registered and active in Russia in 2011--2023 from the legally binding administrative data. The information includes primary industry code, address of incorporation, legal form, etc. With this data we define the entities that are mandated to file their financial statements and append them as having missing financials in the RFSD. This way, we are able to understand the magnitude and sign of the reporting bias: are the eligible non-filers systematically different from the filing firms? We observe an alarming pattern: only 44.1\% of the expected annual filings by eligible firms in 2011--2023 are present in the administrative data underlying the RFSD. While we are able to reconstruct additional 5.5\% of missing filings from next-year statements, failure to file is an important source of selection bias. Firms designated by the government as strategic or firms exiting the market tend not to file. Non-filing is also found to exhibit serial correlation. In contrast, state-owned firms show better filing discipline. To the best of our knowledge, the RFSD is the only openly available country-level dataset with financial statements that includes non-filing firms and thereby allows for explicit handling of non-reporting and selection bias.

\section*{Methods}

\begin{figure}
    \centering
    \includegraphics[width=1\linewidth]{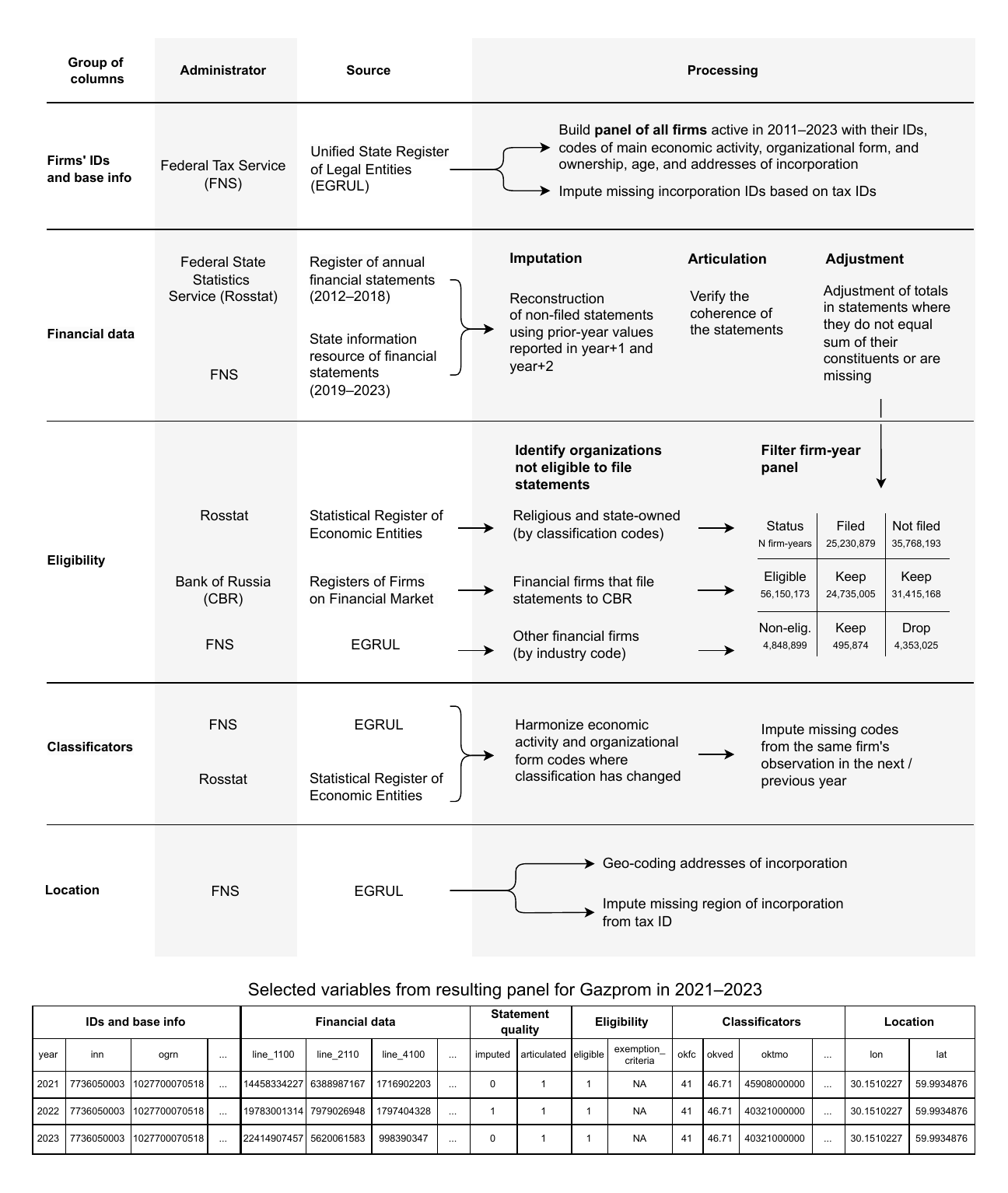}
    \caption{Schematic overview of the RFSD construction}
    \label{fig:panel_construction}
\end{figure}

The process to construct the RFSD is outlined in Figure~\ref{fig:panel_construction}. Below we provide a step-by-step overview.

\paragraph{Acquiring information on the universe of firms}
The Federal Tax Service of Russia (FNS) administers the Uniform State Register of Legal Entities (EGRUL)\cite{egrul}. It contains the official and legally binding information on every active organization in the country. We purchased access to this resource from the FNS and gathered the end-of-year snapshots for 2015--2023 that also included organizations that had been dissolved by 2015. Each snapshot is a collection of millions of eXtensible Markup Language (XML) files (one per firm) with basic information such as firm name, taxpayer identifier, address of incorporation, main and secondary NACE Rev.~2-compatible industry codes, organizational form (stock corporation, limited liability company, government agency, etc.), date of incorporation or liquidation. We developed the parsers for robust extraction of this information from individual XML files and stored it in a flat table with 60,999,072 firm-year observations for the period 2011--2023 where each firm is uniquely identified by a combination of its taxpayer identifier (INN) or organization identifier (OGRN).%(56150173 + 4848899)

\paragraph{Defining eligible firms}
Most Russian organizations are required to file annual unconsolidated financial statements with the Federal State Statistics Service of Russia (Rosstat) before 2019 and with the FNS since 2019. The following entities are exempt from this requirement by law: government bodies and government-owned public service providers such as schools or hospitals, religious organizations, and financial organizations that submit statements to the Central Bank of Russia (CBR), such as banks, insurance companies, brokers\cite{p4art18fz402}. Finally, the firms incorporated in the last quarter of the year are not required to file their statements for that year\cite{p3art15fz402}. The Rosstat maintains the Statistical Register of Economic Entities with the information on government ownership and organizational form for all Russian organizations. We extend the information from the EGRUL with these codes and define government and religious entities based on their year-varying organizational form and/or ownership codes. Financial firms are defined following the CBR's Registers of Professional Participants of Financial Market\cite{cbrprofregisters} supplemented with the list of credit, insurance institutions and investment, private pension funds from the Register of Firms on Financial Market\cite{cbrregister}. Newly incorporated firms are defined based on its quarter of incorporation. These rule-based exclusions produce 4,848,899 ineligible and 56,150,173 eligible firm-years from the universe of 60,999,072 organization-years active in 2011--2023. Ideally, we would expect all eligible firms to file their annual statements.

% The Methods should include detailed text describing any steps or procedures used in producing the data, including full descriptions of the experimental design, data acquisition assays, and any computational processing (e.g. normalization, image feature extraction). See the detailed section in our submission guidelines for advice on writing a transparent and reproducible methods section. Related methods should be grouped under corresponding subheadings where possible, and methods should be described in enough detail to allow other researchers to interpret and repeat, if required, the full study. Specific data outputs should be explicitly referenced via data citation (see Data Records and Citing Data, below).

% Authors should cite previous descriptions of the methods under use, but ideally the method descriptions should be complete enough for others to understand and reproduce the methods and processing steps without referring to associated publications. There is no limit to the length of the Methods section. Subheadings should not be numbered.

\paragraph{Acquiring financial statements}
We collect financial statements from the Rosstat (for 2012--2018) and the FNS (for 2019--2023). The statements filed to the Rosstat are publicly accessible in a tabular Comma-Separated Value (CSV) format (one CSV per year comprising statements from all firms) on its website\cite{rosstatdata}. The FNS statements for individual firms are accessible via a fee-based Application Programming Interface (API) called GIR BO\cite{girbo}. We downloaded the Rosstat's yearly CSV files and purchased access to the FNS API to query millions of firm-year XMLs with statements for every active firm. It is also noteworthy that we have maintained a record of non-filing firms, that is to say, active firms (in accordance with the EGRUL) that have no financial statements for the corresponding year in the FNS API. Our query was made in late 2024, which was beyond the filing deadline for 2023 (April 1, 2024), and was designed to make multiple requests for firms that the API returned no statements for to minimise data loss. We believe that this way we captured the universe of statements filed by the Russian firms. Then we developed the procedures for robust extraction of information from the two data providers ---- the Rosstat (CSVs, before 2019) or the FNS (XMLs, from 2019) and formed a flat table with the firm-level financial statements for 2012--2023 where each firm is uniquely identified by its taxpayer identifier (INN). We removed statements for firm-years that had no match in the EGRUL panel of active firms. Our procedures account for the multiple filings per firm-year that occur when firms submit adjusted statements by taking the most recent filing available. Between 160 and 320 thousand firms filed adjusted statements each year in 2019--2023, with some of them revising their statements multiple times.

\paragraph{Imputation}
Russian firms are required to state not only the current year financials but also the preceding year financials (in case of the balance sheet variables, two prior year financials) each year. This fact is leveraged to impute missing statements for the firms not filing in year $t$ but filing in years $t+1$ or $t+2$ (for the balance sheet). This imputation was beneficial as it allowed us to reconstruct additional 3,060,732 statements. This approach proved particularly fruitful in the case of 2011, for which no individual statement could be identified within the Rosstat dataset. Figure~\ref{fig:validation_plots}(d) shows that the imputation allows us to restore data for about 5\% firms each year. Figure~\ref{fig:validation_plots}(b), in turn, reports the contribution of the restored revenue and materials to the yearly totals in the RFSD. It is important to note that we only impute the entirely missing statements. If a value$_t$ in statement$_t$ differs from value$_{t-1}$ reported in statement$_{t+1}$ it may be due either to a mistake correction or to a change in reporting standards. As it is impossible to distinguish between the two automatically, we leave the not-missing statements as is and do not use the information from $t+1$ or $t+2$ to correct year-$t$ statements. Another thing to note is that a statement reconstruction done in the manner described could not be full in Rosstat's years as the CSV files it provided did not feature prior-year values for the entire cash flow statement and some other variables. Finally, financial statements provided by the Rosstat and the FNS differ significantly in the way of handling missing values. Where the FNS provides XML files that simply lack a field if a firm has not filled it, the Rosstat's CSV files have zeros in place of missing values. In the latter case it is impossible to distinguish between truly zero values submitted by a firm and absent values. Consequently, in the 2011--2018 period, all zeros were treated as missing data.

\paragraph{Harmonization}
The classification of organizational and legal forms (OKOPF) underwent a change in 2013, while classification of industry codes was modified in 2014. We use the official correspondence tables to harmonize the said codes across 2011--2023. In the case of some firm-years, classification codes were absent; in these instances, the codes were imputed on the basis of the same firm's observation in the next or previous year. Furthermore, we harmonized financial statements across the data providers and units of measurement (rubles, thousands of rubles, or millions). We also harmonize the statements across the two report forms available to the firms. Small- and medium-sized enterprises in Russia have an option to file simplified statements with less detailed balance sheet and profit and loss statement and without the cash flow statement\cite{simplified}. Statement forms remained stable throughout the covered period. However, in 2019 full form of the profit and loss statement was changed\cite{changeform}. The change concerned three tax variables (current income tax, income tax adjusted on deferred tax assets and liabilities, and deferred tax assets and liabilities) that were consolidated in one variable. The change was effective for accounts filed starting from 2020, but filers were allowed to use the new form in 2019 filings as well. The data does not allow us to distinguish between firms using old and new forms in 2019 and we do not consolidate the tax lines in that year.

\paragraph{Adjustment of totals}
The financial statements comprise a set of variables, which provide a summation (totals) of certain sections. These include non-current assets (line 1100), current assets (line 1200), total assets (line 1600), etc. It is  necessary to verify that the values displayed in such totals are equal to the sum of their respective components. To illustrate, line 1100 must be equal to the sum of lines 1110, 1120, and so on up to line 1190. Similarly, line 1600 must be equal to the sum of lines 1100 and 1200. In the event that the discrepancy between the stated total value and the calculated one exceeds 4~thousand rubles (the threshold for this discrepancy is derived from the FNS recommendations for statement articulation verification\cite{fnsrefvalues}), or is absent, the calculated value is substituted in its place. Additionally, we add the calculated total lines that are not included in the simplified statements form to ensure their compatibility with the full statements.

\paragraph{Data scope}
We have the universe of 60,999,072 organization-years active in 2011--2023 from the EGRUL, out of which we defined 56,150,173 eligible firm-years. We collected 25,230,879 firm-year financial statement filings from the Rosstat or the FNS for the corresponding period and matched them with the universe of firms on the taxpayer identifier (INN) and year. The lion's share, i.e. 24,735,005 of firm-year filings, comes from the eligible firms. A further 495,874 filings are contributed by the firms we defined as non-eligible to file the statements. This is due to errors on the part of the firms (e.g. small government or religious agencies mistakenly filing their statements), errors in classification codes (when an organization is erroneously defined as government based on its organization code), or changes in the exemption requirements (e.g. some financial firms reporting to the Rosstat instead of the Central Bank in certain years). Importantly, we also have the information on eligible non-filers, or 31,415,168 firm-years that we deemed as eligible but who failed to submit their statements. Finally, as expected, we detect 4,353,025 non-eligible firm-years that have no filings. We exclude such non-eligible non-filers from the data set altogether. These are mostly government agencies, religious, or financial entities who are not required to file their accounts to the Rosstat or the FNS. We arrive at a panel with the universe of the eligible firms (regardless of their filing status) as well as the non-eligible filers in 2011--2023 with a total of 56,646,047 firm-year observations for 9,560,262 firms.

\paragraph{Data enrichment}
We conducted location inference using the addresses of incorporation reported in the EGRUL for every firm-year since 2014 (end-year of our last EGRUL snapshot). The fact that the addresses are stored in a structured form, with separate fields for region, city (or village), street, and house names or numbers, provides a significant advantage. We set up a local instance with OpenStreetMap Nominatim v. 4.4.0\cite{nominatim}, a fast, scalable, and open geocoding solution, using a Docker container provided by \url{https://github.com/mediagis/nominatim-docker}. Then we performed structured queries to Nominatim to geocode every unique address of incorporation in the EGRUL. The initial query included all the constituent elements of the address, including the region, city, street, and house number. In the event of unsuccessful geocoding, the house was subsequently excluded from the second query.  In the event of a failure, a third query was conducted, utilising solely the region and city names. Subsequently, the obtained geographic coordinates were divided into three categories based on their Nominatim Address Rank. The Address Rank is a value that ranges from 4 to 30 and can be converted back into a specific component of a structured address (for example, 4 represents a country and 30 represents a house). Addresses with a rank of 30 were considered to have been geocoded up to the level of the house, while addresses with a rank between 26 and 29 were treated as having been geocoded up to the level of the street. Addresses with a rank between 12 and 25 were regarded as having been geocoded at the city level.

\paragraph{Variables}
Consider an extract from the RFSD at the bottom of Figure~\ref{fig:panel_construction} for Russia's largest company, Gazprom, in 2021--2023. Each company in the data is identified by either the taxpayer identifier (INN) or the organization identifier (OGRN), with the former being the preferred option due to its use in the FNS API. The data then includes 187 variables from the financial statements: balance sheet (\texttt{line\_1100}--\texttt{line\_1700}), profit \& loss statement (\texttt{line\_2100}--\texttt{line\_2910}), statement of changes in equity (\texttt{line\_3100}--\texttt{line\_3600}), cash flow statement (\texttt{line\_4100}--\texttt{line\_4490}), and statement on the proper use of funds received (\texttt{line\_6100}--\texttt{line\_6400}). The numbers in variable names correspond to the official codes of the variables\cite{financialforms} such that \texttt{line\_1100} are non-current assets, \texttt{line\_2110} is revenue, and \texttt{line\_4100} is cash flow from operating activities. The full table with English-language names and descriptions of the variables is available in Table~\ref{table:variables} in the Supplementary Materials.
Apart from these regular lines, there are optional (decoding) lines which a firm may use to further detail its statement. Unlike the regular variables, these have no dedicated numbers and are named as the firm sees fit. To illustrate, Gazprom's profit and loss statements detail its revenue in \texttt{line\_2110} by source in the decoding lines: revenue from gas sales, oil sales, petrochemicals sales. Since both the naming and the logic behind decoding is not uniform, it is not feasible to include these optional lines in a flat table (and because of that this information is absent in the Rosstat's CSVs, 2012--2018, and only present in the FNS' XMLs, 2019--2023). However, cash flow statements do not articulate if such lines are present and one does not take them into account. This poses a problem since the decoding lines are widely used by larger firms, present in more than 200,000 observations in 2019--2023 with average revenue close to 2 bln rubles. We parse the optional lines present in cash flow statements~--- to name them we use an \texttt{x} suffix in place of a last digit that a line's name would have if it was a regular line, e.g. \texttt{line\_411x} (in cases where several additional variables were used to detail the same item, we provide the sum of them). We do not do the same for the balance sheet and other parts of a statement (with the exception of changes in equity), since the structure of these are such that optional lines would simply sum up to the value in an item they decode, and inclusion of the sum of them is therefore redundant. We also flag the statements that were missing but that we were able to impute from the next-year filings. Gazprom, in particular, did not file in 2022, and the statements for that year are reconstructed from the 2023 filings. Apart from the identifiers, the basic information from the EGRUL or other registers is added, such as the primary industry or organizational form codes, or firm age. Finally, we have the longitude and latitude of the firm address of incorporation and its level of detail.

\section*{Data Records}
Data records are composed of the dataset and the accompanying GitHub repository. The RFSD is hosted on Hugging Face (\url{https://huggingface.co/datasets/irlspbru/RFSD}) and Zenodo (\url{https://doi.org/10.5281/zenodo.14622209}). The data is stored in a structured, column-oriented, compressed binary format Apache Parquet\cite{vohra2016apache}. The RFSD data is partitioned by year enabling end-users to query only  variables of interest in years of interest without loading the full data into memory. The GitHub repository (\url{https://github.com/irlcode/RFSD}) holds instructions for importing the data in R or Python environment as well as three use cases. For those engaged in macroeconomic research, we present a replication of study on the interest to cost of goods sold ratio of Russian firms by Mogilat et al.\cite{mogilat2024}, which was based on Interfax's SPARK data. For scholars of industrial organization, we replicate the total factor productivity estimation of Kaukin and Zhemkova\cite{kaukin2023allocation}, which employed Moody's Ruslana data. For economic geographers, we offer a novel model-less house-level GDP spatialization that capitalizes on geocoding of firm addresses.

\section*{Technical Validation}

We conduct an internal and external validation of the RFSD. Internally, we check whether the financial variables are logically consistent and sum to the known amounts. Externally, we check whether the aggregates from the RFSD correspond with the aggregates from independent official or academic sources.

\begin{figure}[t]
\centering
\vspace{-3.8cm}
\includegraphics[width=1\textwidth]{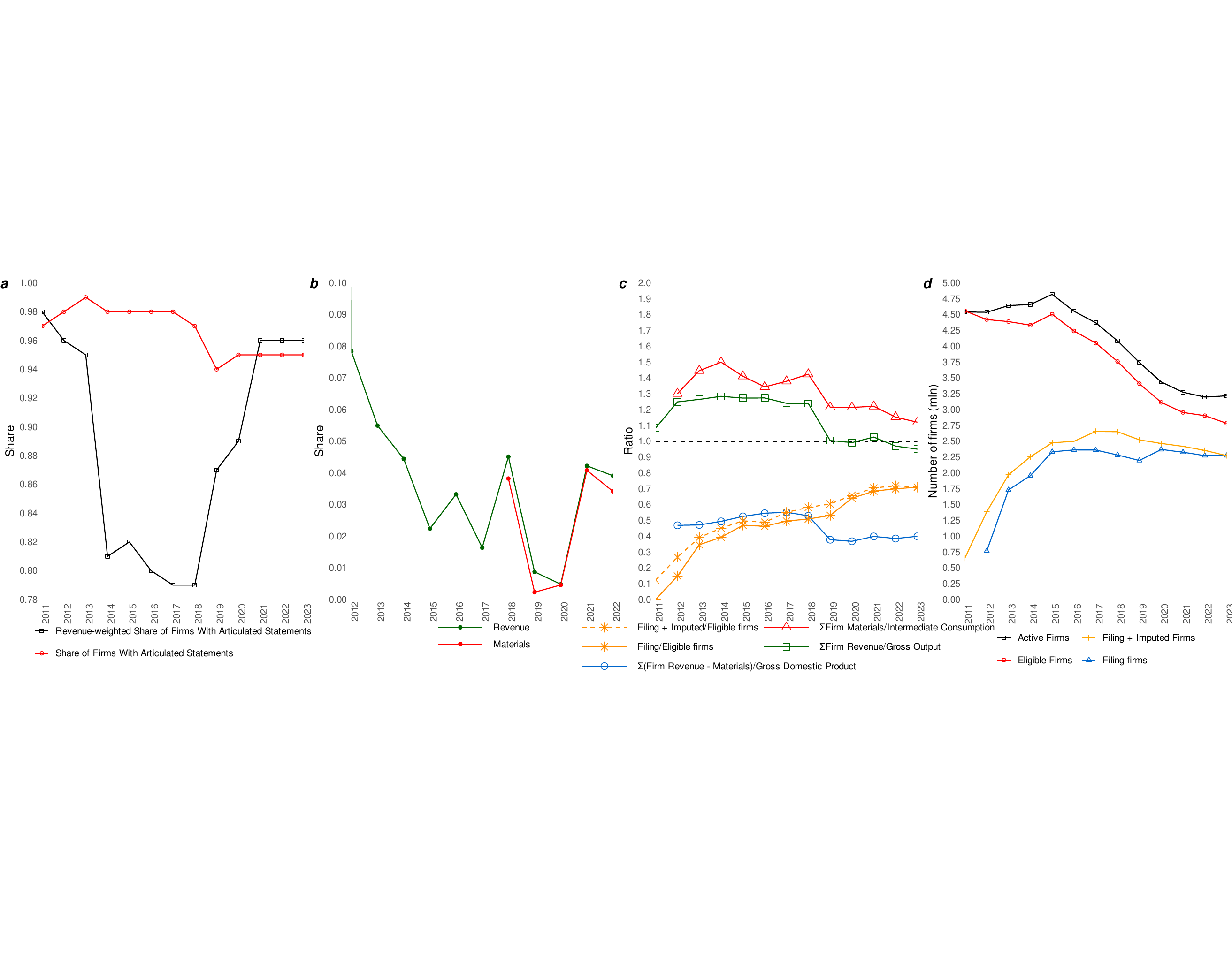}
\vspace{-4.8cm}
\caption{\textbf{Validation of the RFSD}. (\textbf{a--d}). Firms with anomalous values are excluded. (\textbf{a}) Annual shares and revenue-weighted shares of firms with articulated statements. (\textbf{b}) Shares of revenue, materials, and value added imputed from the next-year statements. We define value added as revenue minus materials for firms reporting both non-zero positive variables. (\textbf{c}) Filing rate and Gross Output, Intermediate Consumption, or Value Added in the RFSD vs. National Accounts. (\textbf{d}) Numbers of active firms in the FNS official bulletin on entity registration vs. eligible or filing firms in the RFSD.}
\label{fig:validation_plots}
\end{figure}

\subsection*{Internal validation}
\paragraph{Articulation}
Financial statement articulation is the relationship between its constituent parts that is both logically and mathematically consistent. We perform internal validation of the RFSD by checking whether the individual statements articulate well. In particular, we equate the total values in the summarizing lines (e.g. non-current assets (line 1100), total assets (line 1600), etc.) with the manually calculated totals of their constituent parts. The manual calculation is made following the official guidelines for financial statement articulation by the FNS\cite{fnsrefvalues}. These guidelines define 67 equations for firms submitting full statements and 38 equations for firms filing simplified statements pertaining to different parts of the statement. We use only equations for the balance sheet, profit and loss, and cash flow statements parts~--- 22 equations for full statements and 4 for simplified statements. The full list of equations is available in the Supplementary Materials. We flag a financial statement as articulated if the discrepancy for every applicable equation is within the official threshold of 4~thousand rubles (about USD40 as of late 2024) defined by the FNS\cite{fnsreffour}. We find that only about 5\% of statements do not articulate (see Figure~\ref{fig:validation_plots}(a) for a temporal evolution). We observe that the share of articulating statements was high in the Rosstat data (2011--2018) and decreased slightly following the change in the data provider to FNS (2019--2023). These trends might be misleading, however, because lack of articulation by a large firm might matter more than the erroneous statement filed by a small firm. In light of this, we additionally report revenue-weighted share of articulating statements each year in Figure~\ref{fig:validation_plots}(a). Revenue-weighted articulation plummeted in 2014 and began its gradual increase in the FNS period starting from 2019. We attribute the 2014 drop to changes in accounting rules that tightened the eligibility criteria to submit simplified statements\cite{simpleligible}. This change forced firms to switch to much more detailed full statements that included the decoding lines explained above. The Rosstat data does not handle the decoding lines correctly for many firms, including major companies. The 2019 increase follows the change of the data provider to the FNS and reflects better treatment of the decoding lines in the source data.

\paragraph{Anomalous values}
Internally valid financial statements not only display consistent relationships between their constituent parts, but also report reasonable values. This is especially relevant for the RFSD where we observed two firms reporting revenue that was larger than Gazprom or Rosneft, Russia's largest companies by revenue and capital, by a factor of 8 to 26. Since having such anomalous values is detrimental to any aggregation, we engaged in manual review of top-20 firms in terms of revenue or total assets within each 2-digit industry (excluding financial firms), firms with largest year-on-year changes in key financials, and firms with imputed statements and largest revenues. Our review has identified 436 firms that filed 1,130 anomalous statements in 2011--2023. The judgement was made based on the audit opinions, financials of known industry leaders, firm websites, or public information regarding the firms suspected of reporting anomalous values. We recommend RFSD users to exclude those companies from consideration and do so in our external validation. 

\subsection*{External validation}
\paragraph{Comparison with official aggregates}
To validate the RFSD externally, we leverage Russia's National Accounts. As our numerator we take the annual sum of revenue, materials, or value added (defined as revenue minus materials) for all non-anomalous firms reporting non-zero positive values in the RFDS. As the denominator we use the Gross Output, Intermediate Consumption, and Gross Domestic Product, respectively, from the National Accounts\cite{nationalacc}. If the resulting ratio is close to one this means that firm-level data aggregates well to the National Accounts. We acknowledge that this comparison is inherently flawed as the two sources are fundamentally different: the unconsolidated financial statements are reported according to the Russian accounting rules on book value while the National Accounts are compiled based on the System of National Accounts rules and are valued at market prices. Gross Domestic Product also accounts for shadow economy and non-market production that may be missing in the RFSD. Figure~\ref{fig:validation_plots}(c) reports the resulting ratios. We find that the RFSD aggregates follow the National Accounts of the Russian economy, with Gross Output and Intermediate Consumption ratios exceeding unity in the Rosstat data before 2019 and closer to unity in later periods. This decrease can be explained by non-filing by country's largest firms and by sanctions-related legislation allowing certain companies to not publish their statements\cite{nogirbodecree}. This is evidenced if we look at the filing ratio, that is the ratio of filing to eligible firms, in Figure~\ref{fig:validation_plots}(c). It displays steady increase throughout the years. GDP ratio, in contrast, shows that the value added in the RFSD comprises only 40\%--50\% of the GDP in the National Accounts. This should not be viewed as indication of data deficiencies and is explained by the aforementioned differences in reporting in the System of National Accounts.

Apart from the National Accounts, we validate the RFSD against the FNS statistics. In Figure~\ref{fig:validation_plots}(c) we compare the annual count of active firms from the FNS official bulletin\cite{fnsegruln} with the number of eligible and filing firms in the RFSD. Each year 280,667 firms are deemed as ineligible on average, and this relationship remains stable over time starting from 2013. We attribute small differences between the counts in 2011--2012 to erroneous extraneous observations in our EGRUL data. The number of filing firms increases with time, suggesting better compliance with the filing requirement in later years. Conversely, the number of active or eligible firms displays a steady decrease since 2016. This is due to the effort of the FNS to identify and liquidate inactive or fly-by-night firms established for tax evasion and managerial diversion purposes. Prior to 2016 approximately 32\% of all firms were reportedly identified as rogue, while by the end of 2019 their share had drastically decreased to 3.1\%\cite{mishustin2019}. Finally, Figure~\ref{fig:validation_plots}(d) shows the beneficial effect of our imputation procedure as 235,441 firms have their statements restored from the next-year filings on average each year.

\begin{table}[t]
\begin{adjustbox}{width=1\textwidth}
\centering 
\input{orbis_completeness}
\end{adjustbox}
\caption{\textbf{Comparison of the RFSD with Orbis in 2021.} We consider Russian non-government firms with over \$1 million in revenue in 2021 in the two data sets and report the total, mean, and median value of revenue and assets for firms based on their presence in the data sets. Firms with anomalous values are excluded.}
\label{table:orbis_completeness}
\end{table}

\paragraph{Comparison with Orbis}
Moody's Orbis is the primary source of firm-level information for developed and developing countries\cite{kalemli2024construct,bajgar2020,dall2022using}. Its component for Russian, Ukrainian, and Kazakhstan companies, called Ruslana, is sourced, \textit{inter alia}, from the same administrative data provided by the Rosstat and the FNS that we use to construct the RFSD. It is therefore of value to compare data completeness of the RFSD with that of Orbis, especially in light of its ubiquity in the literature. We queried Orbis to extract all Russian entities with known global ultimate owner and over \$1 million in revenue in 2021, excluding public authorities. Our query was made in April, 2023 and included only active companies with known financial data for 2021\cite{knorre2024stakeholder}. We impose these restrictions on the Orbis sample due to infeasibility of exporting the full list of all Russian organizations. We then match the Orbis sample with the RFSD on taxpayer and organization identifiers and compare the coverage for 2021 in Table~\ref{table:orbis_completeness}. Out of 185,222 firms with financials for 2021 available in our Orbis sample, the vast majority of firms (182,641, 98.6\%) also have their financials in the RFSD for 2021. The firms present in both data sets had over \$2 trillion of total revenue or assets, forming the bulk of the Russian economy in 2021. The RFSD also includes 58,835 firms reporting over \$1 million in revenue in 2021 that are missing in Orbis (second row of Table~\ref{table:orbis_completeness}). These missing firms have mean and median revenue and assets comparable to the present firms. Large number of missing firms in Orbis \textit{vis-à-vis} the RFSD highlights that the latter source has non-trivial amount of additional data not present in Orbis. Finally, 2,581 firms with financials in Orbis did not have financials in the RFSD and were flagged as non-filers. These firms were responsible for a non-trivial amount of total revenue (\$209 billion) and had larger revenue and total assets on average, suggesting non-random data omissions in the RFSD. We manually examined the leading missing firms in terms of revenue and found that their financials were retrospectively excluded from the FNS API. Starting from financial statements for 2018 major Russian firms were authorized not to disclose their financial statements\cite{nogirbodecree}. The list initially included 11 firms\cite{residentslist}, but after 2022 was expanded to cover more than 1,000 firms\cite{putindecree2, nogirbodecree2}. We believe that the observed retroactive redaction of the FNS API data and missing financials in the RFSD are explained by firms exercising their right not to disclose.

\begin{figure}
\centering
\includegraphics[width=\linewidth]{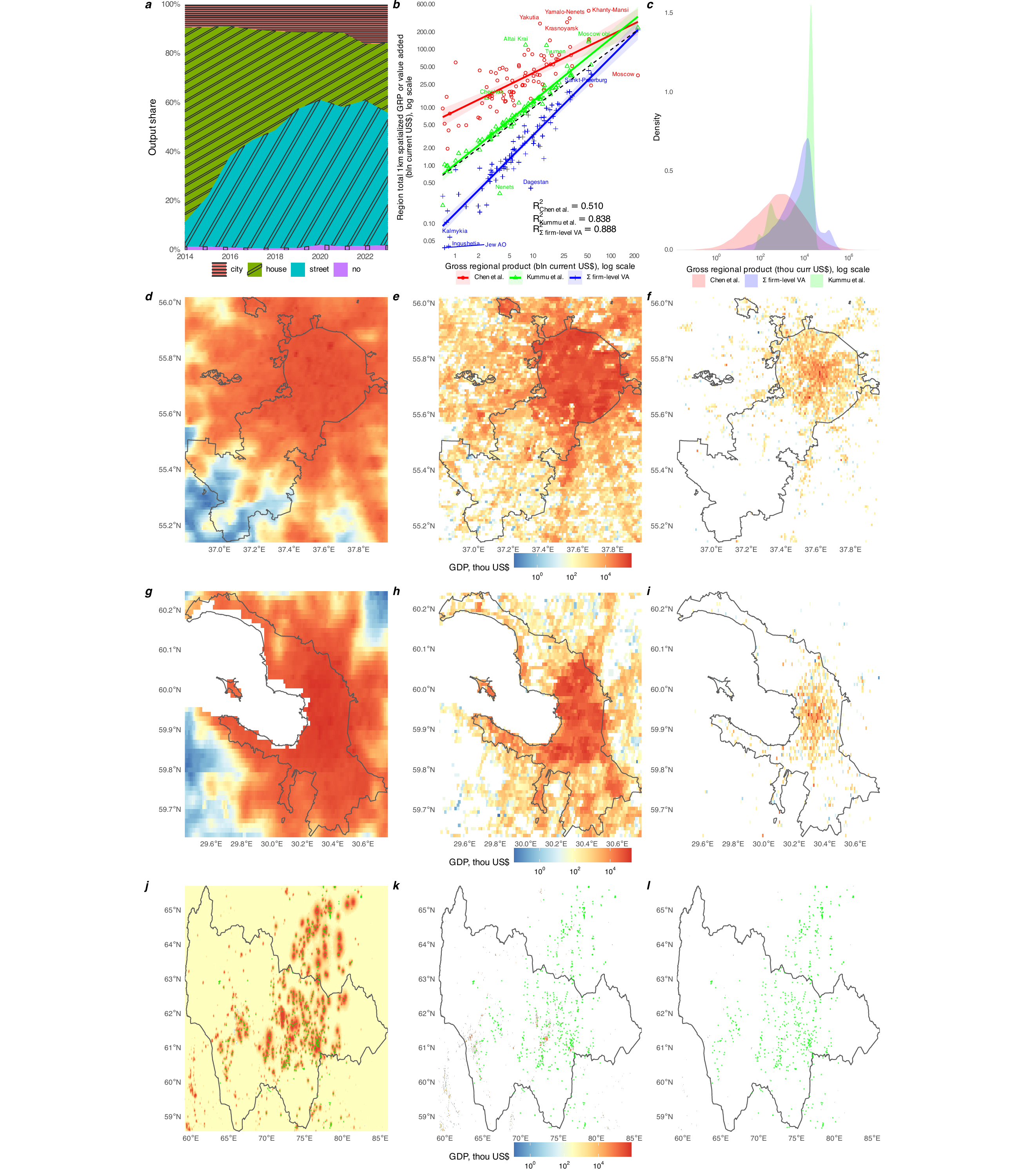}
\vspace{-0.8cm}
\caption{\textbf{Spatial validation}. (\textbf{a--l}). (\textbf{a}) Revenue-weighted share of Russian firms in 2011--2023 by geocoding level of their address of incorporation. (\textbf{b}) Official gross regional product in 2015 and regional totals of its 1~km$\times$1~km spatializations in 2015 from Chen et al.\cite{chen2022global}, Kummu et al.\cite{kummu2018gridded}, or firm-level value added totals for geocoded firms from the RFSD. Coloured solid lines are trends from linear log-log regressions, shaded areas are 95\% CI, R$^2$ from these regression are reported. Black dashed line is ideal alignment. Regions displaying largest absolute difference with the official data are annotated. (\textbf{c}) Kernel density estimates of pixel-level 1~km$\times$1~km GDP in 2015's Russia from Chen et al., Kummu et al., or RFSD firm-level value added on Kummu et al. grid. Non-zero pixels only. (\textbf{d}) Chen et al. non-zero data, 1~km$\times$1~km, Moscow, 2015. (\textbf{e}) Kummu et al. non-zero data, 1~km$\times$1~km, Moscow, 2015. (\textbf{f}) RFSD geolocated firm-level non-zero value added, 1~km$\times$1~km, Moscow, 2015. (\textbf{g}) Chen et al., Saint Petersburg. (\textbf{h}) Kummu, Saint Petersburg. (\textbf{i}) RFSD, Saint Petersburg. (\textbf{j}) Chen et al., Khanty-Mansia. Green dots are gas flare locations in 2015 from World Bank Global Gas Flaring Tracker.\cite{lorenzato2022financing} (\textbf{k}) Kummu et al., Khanty-Mansia. (\textbf{l}) RFSD, Khanty-Mansia. Regional boundaries are due to geoBoundaries\cite{runfola2020geoboundaries}.}
\label{fig:spatialization}
\end{figure}

\paragraph{Spatial comparisons}
Financial statements in the RFSD are enriched with locational information regarding the address of incorporation. Figure~\ref{fig:spatialization}(a) reports the revenue-weighted share of firms by geolocation quality. We find that throughout 2014--2023 88.8\% of total revenue is geocoded up to a house or street on average in the RFSD; location of 10.0\% of revenue is available at city level only, the remaining 1.2\% of revenue is impossible to geolocate. We then proceed to assess the validity of the geocoding by comparing the spatially aggregated value added (defined as revenue minus materials) in the RFSD with widely used sources of 1~km$\times$1~km gridded GDP data from Kummu et al.\cite{kummu2018gridded} or Chen et al.\cite{chen2022global} for Russia in 2015. In Figure~\ref{fig:spatialization}(b) we report region-level aggregates from the three spatializations on the $y$-axis versus the Rosstat's official reported Gross Regional Product for 2015 on the $x$-axis. It is immediately evident that Chen et al. data product is ill-aligned with the official data in Russia. For instance, in 2015 Chen et al. reported mere \$37 billion GDP for Moscow in stark contrast to the official Gross Regional Product (GRP) of \$223 billion (all values henceforth are converted to 2015 nominal USD). Error in the opposite direction is observed for the oil-producing region of Khanty-Mansia, where Chen et al. report GRP of \$474 billion, while the official GRP is only \$52 billion. Extreme upward bias of Chen et al. is further confirmed when we compare regional aggregates with Kummu et al. data product. Regression of log aggregate spatialized GRP on log official GRP reveals that the RFSD spatialization has the highest share of variance explained, with Kummu et al. being a close second, and Chen et al. a distant third. Two things contribute to the large upward bias of Chen et al. spatialization and slight inferiority of Kummy et al. data product in relation to the RFSD. First, it is the mechanical imputation of GDP in uninhabited areas by Chen et al. Consider the raw 1~km$\times$1~km pixels for Moscow or Saint Petersburg in Figures~\ref{fig:spatialization}(d--i) for Chen et al., Kumu et al., and the RFSD spatializations, respectively, for 2015. Chen et al. spatialization reports non-zero economic activity in almost every land pixel, while Kummu et al. and RFSD feature much more pixels with zero GDP. Additionally, the RFSD produces much more focused locations with non-zero value added as suggested by ~\ref{fig:spatialization}(c) with density of non-zero pixels in the three data sources in Russia in 2015. This should come as no surprise since the RFSD relies on addresses of incorporation which are located in the settled areas. Second, inadequate handling of gas flares~--- combustion systems utilized in oil wells to incinerate flammable gases, predominantly methane, that are released during the oil extraction process,~--- by Chen et al. also contributes to the upward bias of their data product. Consider the raw pixels for Khanty-Mansia from the three data sources in Figure~\ref{fig:spatialization}(j--l), with gas flaring locations identified by the World Bank\cite{lorenzato2022financing} for 2015 superimposed as dots. Most economic activity in Chen et al. spatialization in Khanty-Mansia is misrepresented as being situated in the sites where gas flaring are also observed. Given that Chen et al. data is ultimately based on nighttime lights, we view lack of gas flare filtering as a serious drawback of this resource. In contrast, both Kummu et al. data product and the RFSD are free of the gas flaring bias.

\subsection*{Reporting bias}

\begin{table}[t]
% \begin{adjustbox}{width=1\textwidth}
\centering 
\input{models_table}

% \end{adjustbox}
\caption{\textbf{Reporting bias}. Coefficients from ordinary least squares regressions of statement filing, articulation, or anomalous values in the RFSD for 2012--2023 on firm characteristics, excluding firms in their first year. All specifications include year, 2-digit primary industry code, and region of incorporation fixed effects. Huber-Eicker-White standard errors clustered at the region of incorporation are in parentheses. Stars show significance: $p <.001$ ***, $p <.01$ **, $p <.05$ *.}
\label{table:filing_regression}
\end{table}

We observe gradual increase in the proportion of eligible firms filing their financial statements between 2012 and 2023 in Figure~\ref{fig:validation_plots}(c). However, the average filing rate is still only 71.2\% in 2023, meaning that almost one third of eligible firms neglected their duty to file. The extant literature does not reach a consensus on whether private companies tend to report statements of lower quality,\cite{habib2018, beuselinck2023private, ball2005} although corporate governance studies of Russian firms support the notion that small\cite{braguinsky2015} and private companies\cite{bagaeva2008a, goncharov2005} tend to under-report their financials. Non-filers may comprise abandoned or fly-by-night firms contributing to the shadow economy. In addition, we observe regional disparities in reporting, with Ingushetia, Chechnya, and Dagestan demonstrating a substantially lower filing rate (49.6\%, 48.7\%, 43\% respectively) than the national average in 2023. Firms in the capital city also are less likely to file: only 62.2\% firms in Moscow filed in 2023. 

\paragraph{Covariates of reporting}
We estimate linear probability models, regressing filing by eligible companies, statement articulation if filed, and reporting anomalous values if filed on selected firm characteristics in 2012--2023. $Strategic_{i,t}$ indicates whether an $i$-th firm is on the list of strategic companies\cite{strategiclistgov, strategiclistputin} authorized not to disclose their financial statements for year $t$. By the end of 2023 the list of strategic companies included 1,133 firms. $Sanctioned_{i,t}$ indicates if a firm in under sanctions imposed by the international community. This variable is constructed by matching the time-varying lists of sanctioned entities from OpenSanctions, an international database of persons and companies of political, criminal, or economic interest,\cite{opensanctions} with the RFSD on taxpayer or organization identifiers or firm names. In case of matching on firm names we performed a manual match to ensure that we flag the correct entities as being under sanctions. We consider only the sanctions ever imposed on 3,643 firms by the Group of Seven Countries, Australia, New Zealand, and Switzerland as the most consequential ones. Both sanctioned and strategic companies are authorised not to disclose their financial statements\cite{nogirbodecree2}, but the two characteristics do not overlap: 55.7\% of firms designated as strategic were not under sanctions. To better understand the interaction between the two, we include an additional $Strategic_{i,t}$ $\times$ $Sanctioned_{i,t}$ term to flag strategic firms that are also under sanctions. $Exit_{i,t}$ indicates whether a firm is liquidated on any ground, that is, it exits a market in year $t$. We include this variable to capture lack of incentives to report after firm exit. $State{\text -}owned_{i,t}$ indicates whether a firm is directly or partially owned by the state, judging by its classifications codes from the EGRUL or the Statistical Register of Economic Entities. The first and the second models also include lagged dependent variables to explore possible serial correlation in decision-making. In consideration of the presence of lagged variables we exclude newly-incorporated firms in their first year of activity from estimation sample in all models. We are also mindful of the possible multicollinearity between $Strategic_{i,t}$ and $Sanctioned_{i,t}$, given large overlap of the two lists. For this reason we include those two characteristics in stepwise fashion in regression models.

Table~\ref{table:filing_regression} presents the results. Filing is negatively associated with being a strategic firm. Statistically and economically significant negative coefficient for filing by strategic firms of -0.230 (p-value: <0.001) in model (3) confirms systematic under-representation of information by the largest Russian companies in the RFSD that we initially uncovered during comparisons with Orbis. Given that mean filing rate in the sample is 49.4\%, we observe a 53.4\% decrease of filing for stategic firms. Being under sanctions, however, is not directly associated with non-filing in our saturated model (3) in Table~\ref{table:filing_regression} (p-value: 0.0676). Instead, being sanctioned offers a weak mediating effect for strategic firms (p-value: 0.0155). This indicates that small- and medium-sized firms under sanctions but not deemed strategic still have the incentives to file. The absence of incentives to disclose financial information for liquidated firms is our leading explanation for the observed failure to file among the exiting firms, with statistically significant coefficient of -0.313 in model (3) (p-value: <0.001). In contrast, state-owned eligible firms demonstrate a statistically significant higher level of discipline and are more likely to file (coefficient: 0.040 in model (3), p-value: <0.001). Filing also exhibits strong serial correlation (coefficient: 0.6016 in model (3), p-value: <0.001).

Apart from filing, we study articulation of filed statements. In model (6) in Table~\ref{table:filing_regression} we do not uncover a robust relationship between statement quality and being strategic or under sanctions (p-values: 0.373 and 0.040, respectively). State-owned firms, being more disciplined filers, however, tend to produce statements that are slightly less likely to articulate, but this results is not statistically significant (p-value: 0.011). Finally, we consider the filing of anomalous or implausible values for 436 firms we identified. Submitting anomalous statements is found to have little to no association with firm characteristics (apart from being strategic). This supports the notion that anomalous values are due to random errors.

\section*{Code availability}
The code used to build the RFSD is available at the dedicated GitHub repository (\url{https://github.com/irlcode/RFSD}). With access to the fee-based FNS API, it is possible to replicate our procedures to obtain, impute, and harmonize the financial statements data. To replicate the RFSD fully one would also need a panel of all active firms and their classification codes, a geo-coding pipeline, etc., that were built outside of the project and are not documented in the repository.

\bibliography{bibliography}

\section*{Acknowledgements}

We thank Ruslan Kuchakov for contributing to the development of the EGRUL parser.

\section*{Author contributions statement}

S.B.: Software, Data Curation, Writing~--- Original Draft. V.L.: Software, Data Curation, Visualization, Legal and Accounting Rules Research, Writing~--- Original Draft. D.S.: Conceptualization, Methodology, Validation, Writing~--- Review \& Editing.

\section*{Competing interests}
The authors declare that they have no known competing financial interests or personal relationships that could have appeared to influence the work reported in this paper.

\clearpage
\appendix
% Reset table numbering
\setcounter{section}{0}

\setcounter{table}{0}
\setcounter{figure}{0}
\setcounter{footnote}{0} 

\renewcommand{\thetable}{A.\arabic{table}}
\renewcommand{\thefigure}{A.\arabic{figure}}
\renewcommand{\thesubsection}{\Alph{subsection}}

%\pagenumbering{Roman}
%\renewcommand{\thepage}{App.~\arabic{page}}

\renewcommand \thepart{}
\renewcommand \partname{}

\section*{Supplementary Materials}
\label{app}

\begin{tabularx}{1\textwidth}{p{0.1\textwidth}p{0.3\textwidth}p{0.55\textwidth}}
\caption{Definition of variables in the RFSD}\\
\input{variables}
\label{table:variables}
\end{tabularx}

\begin{table}
\caption{Official articulation equations used in this paper}
\begin{tabularx}{1\textwidth}{>{\raggedleft\arraybackslash}p{0.2\textwidth} X p{0.75\textwidth}}
\input{articulation_equations}
\end{tabularx}
\label{table:articulation_equations} 
\end{table}

\end{document}

%% file: orbis_completeness.tex
\begin{tabular}{lllllllll}
\toprule 
\multicolumn{2}{l}{Has financials} & \multirow{2}{*}{N Firms} & \multicolumn{3}{c}{Revenue} & \multicolumn{3}{c}{Total assets}\tabularnewline
\cmidrule(lr){1-2}\cmidrule(lr){4-6}\cmidrule(lr){7-9}
RFSD & Orbis &  & Sum, bln \$ & Mean, thou \$ & Median, thou \$ & Sum, bln \$ & Mean, thou \$ & Median, thou \$\tabularnewline
\cmidrule(lr){1-1}\cmidrule(lr){2-2}\cmidrule(lr){3-3}\cmidrule(lr){4-4}\cmidrule(lr){5-5}\cmidrule(lr){6-6}\cmidrule(lr){7-7}\cmidrule(lr){8-8}\cmidrule(lr){9-9}
Yes & Yes & 182,641 & 2,144 & 11,742 & 2,212 & 2,077 & 11,375 & 1,153\tabularnewline
Yes & No & 58,835 & 802 & 13,643 & 2,328 & 855 & 14,533 & 1,345\tabularnewline
No & Yes & 2,581 & 209 & 81,091 & 7,682 & 290 & 112,439 & 6,969\tabularnewline
\bottomrule
\end{tabular}

%% file: models_table.tex
\resizebox{17cm}{!}{
\begingroup
\centering
\begin{tabular}{lccccccccc}
\midrule
Model & (1) & (2) & (3) & (4) & (5) & (6) & (7) & (8) & (9) \\  
& \multicolumn{3}{c}{Filed$_{i,t}$} & \multicolumn{3}{c}{Articulated$_{i,t}$} & \multicolumn{3}{c}{Anomalous$_{i,t}$} \\ \cmidrule(lr){2-4} \cmidrule(lr){5-7} \cmidrule(lr){8-10}

Filed$_{i,t-1}$ & 0.6016$^{***}$ & 0.6017$^{***}$ & 0.6016$^{***}$ & & & & & & \\   
& {\footnotesize (0.0054)} & {\footnotesize (0.0054)} & {\footnotesize (0.0054)} & & & & & & \\
Articulated$_{i,t-1}$ & & & & 0.6022$^{***}$ & 0.6022$^{***}$ & 0.6022$^{***}$ & & & \\   
& & & & {\footnotesize (0.0132)} & {\footnotesize (0.0132)} & {\footnotesize (0.0132)} & & & \\
Strategic$_{i,t}$ & -0.2647$^{***}$ & & -0.2304$^{***}$ & 0.0067 & & 0.0093 & $-7.28\times 10^{-5}$$^{***}$ & & $-7.26\times 10^{-5}$$^{***}$ \\ 
& {\footnotesize (0.0154)} & & {\footnotesize (0.0193)} & {\footnotesize (0.0116)} & & {\footnotesize (0.0104)} & {\footnotesize ($1.97\times 10^{-5}$)} & & {\footnotesize ($2.04\times 10^{-5}$)} \\
Sanctioned$_{i,t}$ & & -0.0488$^{**}$ & -0.0224$^{.}$ & & -0.0241$^{*}$ & -0.0240$^{*}$ & & $-8.25\times 10^{-5}$$^{*}$ & $-8.25\times 10^{-5}$$^{*}$ \\    
& & {\footnotesize (0.0162)} & {\footnotesize (0.0121)} & & {\footnotesize (0.0114)} & {\footnotesize (0.0115)} & & {\footnotesize ($3.42\times 10^{-5}$)} & {\footnotesize ($3.43\times 10^{-5}$)} \\ 
Strategic$_{i,t}$ $\times$ Sanctioned$_{i,t}$ & & & -0.0576$^{*}$ & & & -0.0178 & & & $7.63\times 10^{-5}$ \\    
& & & {\footnotesize (0.0233)} & & & {\footnotesize (0.0713)} & & & {\footnotesize ($5.71\times 10^{-5}$)} \\ 
Exit$_{i,t}$ & -0.3127$^{***}$ & -0.3127$^{***}$ & -0.3127$^{***}$ & 0.0205$^{***}$ & 0.0205$^{***}$ & 0.0205$^{***}$ & $-1.61\times 10^{-5}$ & $-1.61\times 10^{-5}$ & $-1.61\times 10^{-5}$ \\  
& {\footnotesize (0.0240)} & {\footnotesize (0.0240)} & {\footnotesize (0.0240)} & {\footnotesize (0.0014)} & {\footnotesize (0.0014)} & {\footnotesize (0.0014)} & {\footnotesize ($1.37\times 10^{-5}$)} & {\footnotesize ($1.37\times 10^{-5}$)} & {\footnotesize ($1.37\times 10^{-5}$)} \\ 
State-owned$_{i,t}$ & 0.0395$^{***}$ & 0.0390$^{***}$ & 0.0395$^{***}$ & -0.0134$^{*}$ & -0.0133$^{*}$ & -0.0133$^{*}$ & $9.91\times 10^{-7}$ & $1.02\times 10^{-6}$ & $1.06\times 10^{-6}$ \\ 
& {\footnotesize (0.0060)} & {\footnotesize (0.0061)} & {\footnotesize (0.0060)} & {\footnotesize (0.0052)} & {\footnotesize (0.0052)} & {\footnotesize (0.0052)} & {\footnotesize ($2.58\times 10^{-5}$)} & {\footnotesize ($2.58\times 10^{-5}$)} & {\footnotesize ($2.58\times 10^{-5}$)} \\    
\\
%& \multicolumn{9}{c}{\textsc{Summary statistics}} \\
Sample & \multicolumn{3}{c}{\textsc{Eligible firms aged $>$1}} & \multicolumn{3}{c}{\textsc{Filing firms aged $>$1}} & \multicolumn{3}{c}{\textsc{Filing firms aged $>$1}} \\

Dep. var. mean & \multicolumn{3}{c}{\textsc{0.494}} & \multicolumn{3}{c}{\textsc{0.962}} & \multicolumn{3}{c}{\textsc{$3.61\times 10^{-5}$}} \\   
N (firm-years) & \multicolumn{3}{c}{\textsc{43,048,385}} & \multicolumn{3}{c}{\textsc{22,727,193}} & \multicolumn{3}{c}{\textsc{22,727,193}} \\
Year FE & Yes & Yes & Yes & Yes & Yes & Yes & Yes & Yes & Yes \\   
Industry FE & Yes & Yes & Yes & Yes & Yes & Yes & Yes & Yes & Yes \\   
Region FE & Yes & Yes & Yes & Yes & Yes & Yes & Yes & Yes & Yes \\ 
R$^2$ & 0.501 & 0.501 & 0.501 & 0.373 & 0.373 & 0.373 & 0.000 & 0.000 & 0.000 \\  
\midrule
\end{tabular}
\par\endgroup
}

%% file: variables.tex
\textbf{Variable} & \textbf{Suggested name} & \textbf{Description} \\
\cmidrule(lr){1-3} \vspace{3pt} \\
\multicolumn{3}{c}{\textsc{Firm base info}} \\
\texttt{year} &  & Reporting period \\
\texttt{inn} & & Taxpayer identifier (INN)\\
\texttt{ogrn} & & Organization identifier (OGRN)\\
\texttt{region} & & Region of incorporation\\
\texttt{region\_taxcode} & & Tax code of the region of incorporation\\
\texttt{creation\_date} &  & Date of a firm's registration in EGRUL\\
\texttt{dissolution\_date} & & Date of a firm's exit from EGRUL\\
\texttt{age} & & Firm's age in years in reporting period\\
\vspace{3pt}\\
\multicolumn{3}{c}{\textsc{Eligibility}}\\
\texttt{eligible} & & If a firm was eligible to file a financial statement in reporting period\\
\texttt{exempt\_criteria} & & Criteria of exemption from the obligation to file financial statement\\
\texttt{financial} & & Firm is classified as a financial firm \\
\vspace{3pt}\\
\multicolumn{3}{c}{\textsc{Statement}}\\
\texttt{filed} & & If a firm filed a statement for reporting period\\
\texttt{imputed} & & If a statement was not filed but was (partially) reconstructed from the prior-years values reported in the next statement or the one after that\\
\texttt{simplified} & & If a firm filed a statement using simplified (abbreviated) form\\
\texttt{articulated} & & If values in a statement sum up to respective summarizing lines' values\\
\texttt{totals\_adjustment} & & If summarizing lines' values were missing or did not equate the sums of lines they summarized, and were therefore adjusted\\
\vspace{3pt}\\
\multicolumn{3}{c}{\textsc{Classification codes}}\\
\texttt{okved} & & A firm's industry code in terms of the Russian national classifier of economic activities (OKVED, NACE Rev.2-compatible)\\
\texttt{okved\_section} & & A firm's industry section in terms of the Russian national classifier of economic activities \\
\texttt{okpo} & & A firm's type in terms of the Russian National Classifier of Enterprises and Organizations (OKPO)\\
\texttt{okopf} & & A firm's legal form in terms of the Russian national classifier of organizational and legal forms (OKOPF)\\
\texttt{okogu} & & An organization's type in terms of Russian national classifier of state authorities and administration (OKOGU)\\
\texttt{okfc} & & A firm's ownership form in terms of the Russian national classifier of forms of ownership\\
\texttt{oktmo} & & Code of municipal formation of a firm's incorporation in the Russian national classifier of municipal formations (OKTMO)\\
\vspace{3pt}\\
\multicolumn{3}{c}{\textsc{Location}}\\
\texttt{lon} & & Longitude of a firm's address of incorporation\\
\texttt{lat} & & Latitude of a firm's address of incorporation\\
\texttt{geocoding\_quality} & & Geocoding quality in terms of Nominatim Address Rank\\
\vspace{3pt}\\
\multicolumn{3}{c}{\textsc{Balance Sheet}} \\
\texttt{line\_1100} & \texttt{B\_noncurrent\_assets} & Total non-current assets \\
\texttt{line\_1110} & \texttt{B\_intangible\_assets} & Intangible assets \\
\texttt{line\_1120} & \texttt{B\_research\_development} & Research and development results \\
\texttt{line\_1130} & \texttt{B\_intangible\_exploration} & Intangible exploration assets \\
\texttt{line\_1140} & \texttt{B\_tangible\_exploration} & Tangible exploration assets \\
\texttt{line\_1150} & \texttt{B\_fixed\_assets} & Fixed assets \\
\texttt{line\_1160} & \texttt{B\_tangible\_invest} & Income investments in tangible assets \\
\texttt{line\_1170} & \texttt{B\_fin\_invest} & Financial investments \\
\texttt{line\_1180} & \texttt{B\_def\_tax\_assets} & Deferred tax assets \\
\texttt{line\_1190} & \texttt{B\_other\_noncurrent\_assets} & Other non-current assets \\

\texttt{line\_1200} & \texttt{B\_current\_assets} & Current assets \\
\texttt{line\_1210} & \texttt{B\_inventories} & Inventories \\
\texttt{line\_1220} & \texttt{B\_vat\_receivable} & Value-added tax on acquired assets \\
\texttt{line\_1230} & \texttt{B\_accounts\_receivable} & Accounts receivable \\
\texttt{line\_1240} & \texttt{B\_fin\_invest} & Financial investments \\
\texttt{line\_1250} & \texttt{B\_cash\_equivalents} & Cash and cash equivalents \\
\texttt{line\_1260} & \texttt{B\_other\_current} & Other current assets \\

\texttt{line\_1300} & \texttt{B\_total\_equity} & Total equity \\
\texttt{line\_1310} & \texttt{B\_charter\_capital} & Charter capital (contributed capital, statutory fund, partners' contributions) \\
\texttt{line\_1320} & \texttt{B\_treasury\_shares} & Treasury shares (repurchased from shareholders) \\
\texttt{line\_1340} & \texttt{B\_reval\_assets} & Revaluation of non-current assets \\
\texttt{line\_1350} & \texttt{B\_add\_capital} & Additional capital \\
\texttt{line\_1360} & \texttt{B\_reserve\_capital} & Reserve capital \\
\texttt{line\_1370} & \texttt{B\_retained\_earnings} & Retained earnings (uncovered loss) \\

\texttt{line\_1400} & \texttt{B\_longterm\_liab} & Long-term liabilities \\
\texttt{line\_1410} & \texttt{B\_longterm\_debt} & Long-term borrowings \\
\texttt{line\_1420} & \texttt{B\_def\_tax\_liab} & Deferred tax liabilities \\
\texttt{line\_1430} & \texttt{B\_provision\_liab} & Provisions \\
\texttt{line\_1450} & \texttt{B\_other\_liab} & Other liabilities \\

\texttt{line\_1500} & \texttt{B\_shortterm\_liab} & Short-term liabilities \\
\texttt{line\_1510} & \texttt{B\_shortterm\_debt} & Short-term borrowings \\
\texttt{line\_1520} & \texttt{B\_shortterm\_payables} & Short-term payables \\
\texttt{line\_1530} & \texttt{B\_def\_income} & Deferred income \\
\texttt{line\_1540} & \texttt{B\_provision\_liab} & Provisions \\
\texttt{line\_1550} & \texttt{B\_other\_liab} & Other liabilities \\
\texttt{line\_1600} & \texttt{B\_assets} & Assets \\
\texttt{line\_1700} & \texttt{B\_liab} & Liabilities \\

\vspace{3pt}\\
\multicolumn{3}{c}{\textsc{Profit and Loss Statement} } \\

\texttt{line\_2110} & \texttt{PL\_revenue} & Revenue \\
\texttt{line\_2120} & \texttt{PL\_cost\_of\_sales} & Cost of sales \\
\texttt{line\_2100} & \texttt{PL\_gross\_profit} & Gross profit (loss) \\

\texttt{line\_2210} & \texttt{PL\_commercial\_expenses} & Commercial expenses \\
\texttt{line\_2220} & \texttt{PL\_management\_expenses} & Management expenses \\
\texttt{line\_2200} & \texttt{PL\_profit\_from\_sales} & Profit (loss) from sales \\

\texttt{line\_2310} & \texttt{PL\_income\_participation} & Income from participation in other organizations \\
\texttt{line\_2320} & \texttt{PL\_interest\_receivable} & Interest receivable \\
\texttt{line\_2330} & \texttt{PL\_interest\_payable} & Interest payable \\
\texttt{line\_2340} & \texttt{PL\_other\_income} & Other income \\
\texttt{line\_2350} & \texttt{PL\_other\_expenses} & Other expenses \\
\texttt{line\_2300} & \texttt{PL\_before\_tax} & Profit (loss) before tax \\

\texttt{line\_2410} & \texttt{PL\_income\_tax} & Income tax (Current income tax before 2019-2020) \\
\texttt{line\_2411} & \texttt{PL\_current\_income\_tax} & Current income tax \\
\texttt{line\_2412} & \texttt{PL\_def\_income\_tax} & Deferred income tax \\
\texttt{line\_2421} & \texttt{PL\_tax\_liab} & Permanent tax liabilities (not used after 2019-2020) \\
\texttt{line\_2430} & \texttt{PL\_change\_def\_tax\_liab} & Change in deferred tax liabilities (not used after 2019-2020) \\
\texttt{line\_2450} & \texttt{PL\_change\_def\_tax\_assets} & Change in deferred tax assets (not used after 2019-2020) \\
\texttt{line\_2460} & \texttt{PL\_other\_factors} & Other factors affecting the amount of net profit (fines, etc.) \\
\texttt{line\_2400} & \texttt{PL\_net\_profit} & Net profit (loss) \\

\texttt{line\_2510} & \texttt{PL\_reval} & Result from revaluation of non-current assets which are not included in net profit (loss) \\
\texttt{line\_2520} & \texttt{PL\_other\_operations} & Result from other operations which are not included in net profit (loss)\\
\texttt{line\_2530} & \texttt{PL\_income\_tax\_operations} & Income tax on operations which are not included in net profit (loss)\\
\texttt{line\_2500} & \texttt{PL\_total} & Total financial result for the period \\

\texttt{line\_2900} & \texttt{PL\_basic\_earnings\_share} & Basic earnings (loss) per share \\
\texttt{line\_2910} & \texttt{PL\_diluted\_earnings\_share} & Diluted earnings (loss) per share \\

\vspace{3pt}\\
\multicolumn{3}{c}{\textsc{Statement of changes in equity: Previous reporting period} } \\

\texttt{line\_3100} & \texttt{Epp\_equity} & The size of equity at the end of the year preceding the previous one\\

\texttt{line\_3210} & \texttt{Ep\_incr} & Total equity increase \\
\texttt{line\_3211} & \texttt{Ep\_incr\_net\_profit} & Equity increase due to net profit \\
\texttt{line\_3212} & \texttt{Ep\_incr\_asset\_reval} & Equity increase due to assets revaluation \\
\texttt{line\_3213} & \texttt{Ep\_incr\_income} & Equity increase due to contributions from founders \\
\texttt{line\_3214} & \texttt{Ep\_incr\_add\_share\_issue} & Equity increase due to additional shares issue \\
\texttt{line\_3215} & \texttt{Ep\_incr\_share\_value} & Equity increase due to increase in nominal value of shares \\
\texttt{line\_3216} & \texttt{Ep\_incr\_reorg} & Equity increase due to reorganization \\
\texttt{line\_321x} & \texttt{Ep\_incr\_other} & Other factors of equity increase \\

\texttt{line\_3220} & \texttt{Ep\_decr} & Total equity decrease \\
\texttt{line\_3221} & \texttt{Ep\_decr\_loss} & Equity decrease due to loss \\
\texttt{line\_3222} & \texttt{Ep\_decr\_asset\_reval} & Equity decrease due to assets revaluation \\
\texttt{line\_3223} & \texttt{Ep\_decr\_expenses} & Equity decrease due to expenses\\
\texttt{line\_3224} & \texttt{Ep\_decr\_share\_value} & Equity decrease due to decrease in nominal value of shares \\
\texttt{line\_3225} & \texttt{Ep\_decr\_shares\_number} & Equity decrease due to decrease in number of shares \\
\texttt{line\_3226} & \texttt{Ep\_decr\_reorg} & Equity decrease due to legal entity reorganization \\
\texttt{line\_3227} & \texttt{Ep\_decr\_dividends} & Equity decrease due to payment of dividends\\
\texttt{line\_322x} & \texttt{Ep\_decr\_special} & Other factors of equity decrease listed in optional lines\\

\texttt{line\_3230} & \texttt{Ep\_change\_add} & Change in additional equity \\
\texttt{line\_3240} & \texttt{Ep\_change\_reserve} & Change in reserve equity \\
\texttt{line\_3200} & \texttt{Ep\_equity} & Equity amount as of December 31 of the previous year\\ 

\vspace{3pt}\\
\multicolumn{3}{c}{\textsc{Statement of changes in equity: Current reporting period} } \\

\texttt{line\_3310} & \texttt{E\_incr} & Total equity increase \\
\texttt{line\_3311} & \texttt{E\_incr\_net\_profit} & Equity increase due to net profit \\
\texttt{line\_3312} & \texttt{E\_incr\_asset\_reval} & Equity increase due to assets revaluation \\
\texttt{line\_3313} & \texttt{E\_incr\_income} & Equity increase due to contributions from founders \\
\texttt{line\_3314} & \texttt{E\_incr\_add\_share\_issue} & Equity increase due to additional shares issue \\
\texttt{line\_3315} & \texttt{E\_incr\_share\_value} & Equity increase due to increase in nominal value of shares \\
\texttt{line\_3316} & \texttt{E\_incr\_reorg} & Equity increase due to reorganization \\
\texttt{line\_331x} & \texttt{E\_incr\_other} & Other factors of equity increase \\

\texttt{line\_3320} & \texttt{E\_decr} & Total equity decrease \\
\texttt{line\_3321} & \texttt{E\_decr\_loss} & Equity decrease due to loss \\
\texttt{line\_3322} & \texttt{E\_decr\_asset\_reval} & Equity decrease due to assets revaluation \\
\texttt{line\_3323} & \texttt{E\_decr\_expenses} & Equity decrease due to expenses\\
\texttt{line\_3324} & \texttt{E\_decr\_share\_value} & Equity decrease due to decrease in nominal value of shares \\
\texttt{line\_3325} & \texttt{E\_decr\_shares\_number} & Equity decrease due to decrease in number of shares \\
\texttt{line\_3326} & \texttt{E\_decr\_reorg} & Equity decrease due to legal entity reorganization \\
\texttt{line\_3327} & \texttt{E\_decr\_dividends} & Equity decrease due to payment of dividends\\
\texttt{line\_332x} & \texttt{E\_decr\_special} & Other factors of equity decrease listed in optional lines \\

\texttt{line\_3330} & \texttt{E\_change\_add} & Change in additional equity \\
\texttt{line\_3340} & \texttt{E\_change\_reserve} & Change in reserve equity \\
\texttt{line\_3300} & \texttt{E\_equity} & Equity amount as of December 31 of the reporting year \\ 

\vspace{3pt}\\
\multicolumn{3}{c}{\textsc{Adjustment due to changes in accounting policy and correction of errors}} \\
\texttt{line\_3400} & \texttt{ADJ\_equity\_before} & Total equity Before Adjustments \\
\texttt{line\_3410} & \texttt{ADJ\_policy} & Adjustment Due to Change in Accounting Policy \\
\texttt{line\_3420} & \texttt{ADJ\_error} & Adjustment Due to Correction of Errors After Adjustment \\
\texttt{line\_3500} & \texttt{ADJ\_equity\_after} & Total equity After Adjustments \\

\texttt{line\_3401} & \texttt{ADJ\_undistr\_profit\_before} & Amount of undistributed profit before adjustments \\
\texttt{line\_3411} & \texttt{ADJ\_undistr\_profit\_policy} & Adjustment of the amount of undistributed profit due to changes in accounting policy \\
\texttt{line\_3421} & \texttt{ADJ\_undistr\_profit\_errors} & Adjustment of the amount of undistributed profit due to correction of errors \\
\texttt{line\_3501} & \texttt{ADJ\_undistr\_profit\_after} & Amount of undistributed profit after adjustments \\
\texttt{line\_3402} & \texttt{ADJ\_other\_equity\_before} & Size of other equity items before adjustments \\
\texttt{line\_3412} & \texttt{ADJ\_other\_equity\_policy} & Adjustment of other equity items due to changes in accounting policy \\
\texttt{line\_3422} & \texttt{ADJ\_other\_equity\_errors} & Adjustment of other equity items due to correction of errors \\
\texttt{line\_3502} & \texttt{ADJ\_other\_equity\_after} & Size of other equity items after adjustments \\

\vspace{3pt}\\
\multicolumn{3}{c}{\textsc{Net assets}} \\
\texttt{line\_3600} & \texttt{NA\_net\_assets} & Net assets \\ 

\vspace{3pt}\\
\multicolumn{3}{c}{\textsc{Cash flow statement: Operating activities} } \\

\texttt{line\_4110} & \texttt{CFi\_operating} & Cash inflows from operating activities \\
\texttt{line\_4111} & \texttt{CFi\_sales} & From Sale of Products, Goods, Works, and Services \\
\texttt{line\_4112} & \texttt{CFi\_payments} & From Rental Payments, License Fees, Royalties, Commissions, and Other Similar Payments \\
\texttt{line\_4113} & \texttt{CFi\_resale\_invest} & From Resale of Financial Investments \\
\texttt{line\_411x} & \texttt{CFi\_firm\_specific} & Cash inflows stated in optional lines\\
\texttt{line\_4119} & \texttt{CFi\_other} & Other cash inflows \\

\texttt{line\_4120} & \texttt{CFo\_operating} & Cash outflows from operating activities \\
\texttt{line\_4121} & \texttt{CFo\_materials} & Payments to suppliers (contractors) for raw materials, goods, works, and services \\
\texttt{line\_4122} & \texttt{CFo\_labor} & Labor payments \\
\texttt{line\_4123} & \texttt{CFo\_interest} & Interest on debt obligations \\
\texttt{line\_4124} & \texttt{CFo\_income\_tax} & Corporate income tax \\
\texttt{line\_412x} & \texttt{CFo\_special} & Cash outflows stated in optional lines \\
\texttt{line\_4129} & \texttt{CFo\_other} & Other payments \\

\texttt{line\_4100} & \texttt{CF\_balance\_operating} & Balance of Cash Flows from Operating Activities \\

\vspace{3pt}\\
\multicolumn{3}{c}{\textsc{Cash flow statement: Investing activities} } \\

\texttt{line\_4210} & \texttt{CFi\_invest} & Cash inflows from investments \\
\texttt{line\_4211} & \texttt{CFi\_sale\_noncurrent\_assets} & From Sale of Non-Current Assets (excluding Financial Investments) \\
\texttt{line\_4212} & \texttt{CFi\_sale\_shares} & From Sale of Shares of Other Organizations (Equity Interests) \\
\texttt{line\_4213} & \texttt{CFi\_loan\_repayments} & From Repayment of Loans Granted, From Sale of Debt Securities (Claims for Cash from Other Parties) \\
\texttt{line\_4214} & \texttt{CFi\_dividends\_interest} & From Dividends, Interest on Debt Financial Investments, and Similar Inflows from Equity Participation in Other Organizations \\
\texttt{line\_421x} & \texttt{CFi\_invest\_special} & Cash inflow from investment operations stated in optional lines \\
\texttt{line\_4219} & \texttt{CFi\_invest\_other} & Other inflows from investments \\
\texttt{line\_4220} & \texttt{CFo\_invest} & Cash outflows from investments \\
\texttt{line\_4221} & \texttt{CFo\_acquisition\_assets} & Cash outflows from acquisition, creation, modernization, reconstruction, and preparation for use of non-current assets \\
\texttt{line\_4222} & \texttt{CFo\_acquisition\_shares} & Cash outflows from acquisition of shares of other organizations (equity interests) \\
\texttt{line\_4223} & \texttt{CFo\_acquisition\_debt} & In connection with acquisition of debt securities (claims for cash from other parties), granting loans to other parties \\
\texttt{line\_4224} & \texttt{CFo\_interest\_payments} & Interest on debt obligations included in the cost of investment assets \\
\texttt{line\_422x} & \texttt{CFo\_invest\_special} & Cash outflows from investing listed in optional lines\\
\texttt{line\_4229} & \texttt{CFo\_invest\_other} & Other payments because of investments \\

\texttt{line\_4200} & \texttt{CF\_balance\_invest} & Balance of cash flows from investing activities \\

\vspace{3pt}\\
\multicolumn{3}{c}{\textsc{Cash flow statement: Financial operations} } \\

\texttt{line\_4310} & \texttt{CFi\_fin} & Cash inflows from financial operations\\
\texttt{line\_4311} & \texttt{CFi\_loans} & Receipt of loans and borrowings \\
\texttt{line\_4312} & \texttt{CFi\_owner\_contributions} & Cash contributions from owners (participants) \\
\texttt{line\_4313} & \texttt{CFi\_share\_issuance} & From issuance of shares, increase in ownership interests \\
\texttt{line\_4314} & \texttt{CFi\_bond\_issuance} & From issuance of bonds, promissory notes, and other debt securities \\
\texttt{line\_431x} & \texttt{CFi\_fin\_special} & Cash inflows from financial operations stated in optional lines \\
\texttt{line\_4319} & \texttt{CFi\_fin\_other} & Other inflows from financial operations\\

\texttt{line\_4320} & \texttt{CFo\_fin} & Cash outflows from financial operations\\
\texttt{line\_4321} & \texttt{CFo\_payments\_owners} & To owners (participants) in connection with buyback of shares (ownership interests) or their exit from the organization \\
\texttt{line\_4322} & \texttt{CFo\_payments\_dividends} & For payment of dividends and other profit distribution payments to owners (participants) \\
\texttt{line\_4323} & \texttt{CFo\_debt\_repayments} & In connection with redemption (buyback) of promissory notes and other debt securities, repayment of loans and borrowings \\
\texttt{line\_432x} & \texttt{CFo\_fin\_special} & Cash outflows from financial activities stated in optional lines \\
\texttt{line\_4329} & \texttt{CFo\_fin\_other} & Other cash outflows from financial activities \\

\texttt{line\_4300} & \texttt{CF\_balance\_fin} & Balance of cash flows from financing activities \\

\texttt{line\_4400} & \texttt{CF\_balance} & Balance of cash flows for the reporting period \\

\texttt{line\_4450} & \texttt{C\_balance\_start} & Balance of cash and cash equivalents at the start of the reporting period \\

\texttt{line\_4500} & \texttt{C\_balance\_end} & Ending balance of cash and cash equivalents at the end of the reporting period \\

\texttt{line\_4490} & \texttt{C\_foreign\_currency\_impact} & Impact of foreign currency exchange rate changes relative to the ruble \\

\vspace{3pt}\\
\multicolumn{3}{c}{\textsc{Statement on the proper use of funds received} } \\
\texttt{line\_6100} & \texttt{PU\_start} & Beginning balance of funds at the start of the reporting year \\

\texttt{line\_6210} & \texttt{PU\_entrance} & Entrance fees \\
\texttt{line\_6215} & \texttt{PU\_membership\_fees} & Membership fees \\
\texttt{line\_6220} & \texttt{PU\_designated} & Designated contributions \\
\texttt{line\_6230} & \texttt{PU\_voluntary} & Voluntary property contributions and donations \\
\texttt{line\_6240} & \texttt{PU\_income\_activities} & Profit from income-generating activities of the organization \\
\texttt{line\_6250} & \texttt{PU\_income\_other} & Other \\
\texttt{line\_6200} & \texttt{PU\_total\_received} & Total funds received \\

\texttt{line\_6310} & \texttt{PU\_designated} & Expenses for designated activities \\
\texttt{line\_6311} & \texttt{PU\_aid} & Social and charitable assistance \\
\texttt{line\_6312} & \texttt{PU\_conference} & Expenses for conducting conferences, meetings, seminars \\
\texttt{line\_6313} & \texttt{PU\_other\_events} & Other activities \\

\texttt{line\_6320} & \texttt{PU\_administrative} & Administrative expenses \\
\texttt{line\_6321} & \texttt{PU\_labor} & Labor-related expenses (including accruals) \\
\texttt{line\_6322} & \texttt{PU\_nonlabor} & Payments not related to labor \\
\texttt{line\_6323} & \texttt{PU\_travel} & Expenses for business trips and travel \\
\texttt{line\_6324} & \texttt{PU\_maintenance} & Maintenance of premises, buildings, vehicles, and other property (excluding repairs) \\
\texttt{line\_6325} & \texttt{PU\_repairs} & Repairs of fixed assets and other property \\
\texttt{line\_6326} & \texttt{PU\_other\_administrative} & Other administrative expenses \\

\texttt{line\_6330} & \texttt{PU\_acquisition\_assets} & Acquisition of fixed assets, inventory, and other property \\
\texttt{line\_6350} & \texttt{PU\_other\_expenses} & Other \\
\texttt{line\_6300} & \texttt{PU\_total\_expenses} & Total funds used \\

\texttt{line\_6400} & \texttt{PU\_remaining} & Funds at the end of the reporting period\\

\bottomrule

%% file: articulation_equations.tex
\textbf{Summarizing line number} & & \textbf{Articulation equation} \\
\cmidrule(lr){1-3}\\
\multicolumn{3}{c}{\textsc{Full statements}}\\
% \vspace{1pt}\\
\multicolumn{3}{l}{Balance sheet}\\
\texttt{1100} & $=$ & \texttt{1110} $+$ \texttt{1120} $+$ \texttt{1130} $+$ \texttt{1140} $+$ \texttt{1150} $+$ \texttt{1160} $+$ \texttt{1170} $+$ \texttt{1180} $+$ \texttt{1190}\\
\texttt{1200} & $=$ & \texttt{1210} $+$ \texttt{1220} $+$ \texttt{1230} $+$ \texttt{1240} $+$ \texttt{1250} $+$ \texttt{1260} \\
\texttt{1300} & $=$ & \texttt{1310} $+$ \texttt{1320} $+$ \texttt{1330} $+$ \texttt{1340} $+$ \texttt{1350} $+$ \texttt{1360} $+$ \texttt{1370} \\
\texttt{1400} & $=$ & \texttt{1410} $+$ \texttt{1420} $+$ \texttt{1430} $+$ \texttt{1450} \\
\texttt{1500} & $=$ & \texttt{1510} $+$ \texttt{1520} $+$ \texttt{1530} $+$ \texttt{1540} $+$ \texttt{1550} \\
\texttt{1600} & $=$ & \texttt{1200} $+$ \texttt{1100} \\
\texttt{1600} & $=$ & \texttt{1700} \\
\texttt{1700} & $=$ & \texttt{1300} $+$ \texttt{1400} $+$ \texttt{1500} \\
% \vspace{1pt}\\
\multicolumn{3}{l}{Profit and loss statement}\\
\texttt{2100} & $=$ & \texttt{2110} $-$ \texttt{2120} \\
\texttt{2200} & $=$ & \texttt{2100} $-$ \texttt{2210} $-$ \texttt{2220} \\
\texttt{2300} & $=$ & \texttt{2200} $-$ \texttt{2310} $+$ \texttt{2320} $-$ \texttt{2330} $+$ \texttt{2340} $-$ \texttt{2350} \\
% \vspace{1pt}\\
\multicolumn{3}{l}{Cash flow statement}\\
\texttt{4100} & $=$ & \texttt{4110} $-$ \texttt{4120} \\
\texttt{4110} & $=$ & \texttt{4111} $+$ \texttt{4112} $+$ \texttt{4113} $+$ \texttt{4114} $+$ \texttt{4116} $+$ \texttt{4119} $+$ \texttt{optional decoding lines} \\
\texttt{4120} & $=$ & \texttt{4121} $+$ \texttt{4122} $+$ \texttt{4123} $+$ \texttt{4124} $+$ \texttt{4126} $+$ \texttt{4129} $+$ \texttt{optional decoding lines}  \\
\texttt{4200} & $=$ & \texttt{4210} $-$ \texttt{4220} \\
\texttt{4210} & $=$ & \texttt{4211} $+$ \texttt{4212} $+$ \texttt{4213} $+$ \texttt{4214} $+$ \texttt{4216} $+$ \texttt{4219} $+$ \texttt{optional decoding lines}  \\
\texttt{4220} & $=$ & \texttt{4221} $+$ \texttt{4222} $+$ \texttt{4223} $+$ \texttt{4224} $+$ \texttt{4226} $+$ \texttt{4229} $+$ \texttt{optional decoding lines}  \\
\texttt{4300} & $=$ & \texttt{4310} $-$ \texttt{4320} \\
\texttt{4310} & $=$ & \texttt{4311} $+$ \texttt{4312} $+$ \texttt{4313} $+$ \texttt{4314} $+$ \texttt{4316} $+$ \texttt{4319} $+$ \texttt{optional decoding lines}  \\
\texttt{4320} & $=$ & \texttt{4321} $+$ \texttt{4322} $+$ \texttt{4323} $+$ \texttt{4324} $+$ \texttt{4326} $+$ \texttt{4329} $+$ \texttt{optional decoding lines}  \\
\texttt{4400} & $=$ & \texttt{4100} $+$ \texttt{4200} $+$ \texttt{4300} \\
\texttt{4500} & $=$ & \texttt{4400} $+$ \texttt{4450} $+$ \texttt{4490} \\
\vspace{1pt}\\
\multicolumn{3}{c}{\textsc{Simplified statements}}\\
% \vspace{1pt}\\
\multicolumn{3}{l}{Balance sheet}\\
\texttt{1600} & $=$ & \texttt{1150} $+$ \texttt{1170} $+$ \texttt{1210} $+$ \texttt{1250} $+$ \texttt{1230}\\
\texttt{1600} & $=$ & \texttt{1700}\\
\texttt{1700} & $=$ & \texttt{1300} $+$ \texttt{1410} $+$ \texttt{1450} $+$ \texttt{1510} $+$ \texttt{1520} $+$ \texttt{1550}\\
% \vspace{1pt}\\
\multicolumn{3}{l}{Profit and loss statement}\\
\texttt{2400} & $=$ & \texttt{2110} $-$ \texttt{2120} $-$ \texttt{2330} $+$ \texttt{2340} $-$ \texttt{2350} $-$ \texttt{2410}\\

\bottomrule